\documentclass[twocolumn,aps]{revtex4}
\usepackage{amssymb}
\newcommand{\half}{\frac{1}{2}}
\newcommand{\frth}{\frac{1}{4}}
\newcommand{\egth}{\frac{1}{8}}
\newcommand{\tauv}{\mbox{\boldmath $\tau$}}

\newcommand{\delv}{\mbox{\boldmath $\nabla$}}
\newcommand{\sigv}{\mbox{\boldmath $\sigma$}}
\newcommand{\Astr}{/\!\!\!\!A}

\begin{document}
\title{
Self-interaction and mass in quantum field theory }
\author{R. K. Nesbet}
\affiliation{
IBM Almaden Research Center,
650 Harry Road,
San Jose, CA 95120-6099, USA}
\date{\today}
\begin{center}For {\em PhysRevLett} \end{center}
\begin{abstract}
Qualitative implications of electroweak theory are reconsidered on
the assumption that the unique source of fermion rest mass is 
self-interaction via coupling to gauge fields. 
This implies small but nonzero mass for neutrinos, and suggests that
successive fermion generations are distinct coupled-field
eigenstates of a self-interaction mass operator. 
For a scalar Higgs field, this mechanism can account 
for the SU(2) symmetry breaking of electroweak theory 
without a biquadratic self-interaction.  The implied Higgs 
particle mass could be very small, eluding any search  
limited to heavy particles.
\end{abstract}
 
\maketitle
 
\section{Implications of self-interaction}
In relativistic perturbation theory\cite{FEY49}, the self-interaction
mass of an electron is a sum over momentum transfer, logarithmically 
divergent\cite{WEI39,PAS95} unless the sum is somehow cut off.
This divergence indicates that the theory is incomplete, but its 
logarithmic character implies that any physically valid cutoff
can occur only for very large momentum transfer, or at very small
distances.  A convincing self-contained theory would require a
cutoff mechanism that implies correct observed masses for all fermions.
The implications of standard quantum field theory, modified only by such
a cutoff, are reconsidered here, and appear to resolve several mysteries
remaining in accepted current theory.  This argues for a renewed effort
to understand the cutoff mechanism.
\par A classical point-particle would interact with itself through
the static electromagnetic field generated by its electric charge.
This implies an infinite mass, inconsistent with observed reality.
In quantum field theory the
model is fundamentally different.  Each physical particle is
created from the physical vacuum, which must be postulated to have
no net energy or current density, although polarizable.
The static field of a charged fermion
is compensated due to the net neutrality of all leptons and quarks.
The radiative component remains, generated by virtual excitations of 
gauge fields coupled to each elementary fermion field. 
Any interacting fermion
is necessarily dressed by the accompanying virtual radiation.
Bare fermion mass can simply be omitted from the Lagrangian density,
but self-interaction mass due to virtual radiation fields cannot be
eliminated unless there is no mechanism for virtual excitation. 
An immediate implication is that neutrinos carry a small 
mass, due to virtual excitation of the weak gauge fields.
\par That a physical fermion field is a quasiparticle which acquires 
mass from its self-interaction is consistent with the structure of the
Dirac equation: a mass parameter couples field components of opposite 
parity whose energy values have opposite sign.  The field equation
for a bare fermion coupled to gauge fields can be rewritten so that a
mass parameter replaces a 4-vector transition operator linear in the
self-interaction gauge fields.  The mass parameter takes the form of an
eigenvalue of the transition operator, identified as a self-interaction 
mass operator.  Diagonalization in the Fock space of the interacting
system defines a canonical transformation from bare fermions to dressed 
quasiparticles, identified as physical fermions.  This transformation 
breaks chiral symmetry as it produces nonvanishing mass. Although this
is implicit in standard field theory, an algebraic formulation of the 
theory will be developed here in which this transformation is explicit. 
\par A significant implication of this analysis is that for fermions 
the self-interaction mass operator might very well have several
eigenvalues that correspond to discrete states pushed down below 
successive overlapping continua.  This could explain the existence of 
higher generations of fermions, such as heavy leptons and their
corresponding neutrinos.
\par If local charge neutrality is postulated (cosmic jellium), or
atomic nuclei are treated as point charges, standard renormalization 
theory defines a self-contained quantum electrodynamics\cite{SCH58},
restricted to electrons and the Maxwell field.  Extending Maxwell theory
to nonabelian gauge symmetry\cite{YAM54}, the unified electroweak theory
of Weinberg and Salam\cite{WEI96,CAG98,REN90} incorporates neutrinos,
quarks and SU(2) weak gauge fields.  Fermion masses are given their
observed values, justified by the quantitative success of renormalized 
QED.  Renormalization also justifies absorbing vacuum polarization into
the observed electric charge unit as a renormalized coupling constant.  
\par Assuming that fermion masses are solely an expression of the 
self-interaction induced by coupling to gauge fields might appear to 
conflict with an essential element of electroweak theory.  This theory 
postulates a biquadratic self-interaction from which an assumed scalar
boson (Higgs) field acquires mass. 
This induces a canonical transformation that gives mass to
the weak gauge fields but not to the transformed Maxwell field. 
Yukawa terms that couple the Higgs and fermion fields are independently
postulated, and are assumed to account for fermion mass in general. 
\par
It is argued here that since radiative self-interaction is inherent
in the formalism, these Yukawa terms are unnecessary.  Neutrino mass 
requires at least a right-handed chiral isospin singlet neutrino field, 
probably quantitatively negligible.  Radiative self-interaction
is shown here to be relevant to a scalar boson field.
This can replace the biquadratic Higgs self-interaction\cite{CAW73}
while retaining the essential structure of electroweak theory. 
This argument implies a mechanism that forces the residual mass of the
Maxwell field to zero.  The Higgs mass could result solely from coupling
to the weak gauge fields, and might be very small.  Such a Higgs
particle would be missed by current searches for a heavy scalar boson. 

\section{Self-interaction  mass in QED}
QED theory requires two distinct postulates.  The first is the dynamical
postulate that the action integral $W=\int{\cal L}d^4x$ of the 
Lagrangian density ${\cal L}$ over a specified space-time region is 
stationary with respect to variations of the independent fields $A_\mu$ 
and $\psi$, subject to fixed boundary values.  The second postulate 
attributes algebraic commutation or anticommutation properties 
to these elementary fields.  
\par Defining
$F_{\mu\nu}=\partial _{\mu} A_{\nu}- \partial _{\nu} A _{\mu}$,
the QED Lagrangian density is 
\begin{equation}\label{Eq01}
{\cal L}= -(1/16 \pi ) F^{\mu \nu} F_{\mu \nu} +
 i\hbar c\psi^{\dagger}\gamma^0\gamma^\mu D_\mu \psi .
\end{equation}
Coupling to the electromagnetic 4-potential $A_{\mu}$
occurs through the covariant derivative
\begin{equation}\label{Eq03}
D _{\mu} = \partial_{\mu}+i(-e/{\hbar c})A_{\mu} ,
\end{equation}
where -e is the renormalized electronic charge.
The notation used here defines covariant 4-vectors
\begin{eqnarray*}
x_{\mu} = ( ct,- {\bf r}) , \;
\partial_{\mu} = (\partial/c\partial t,\delv ) , \;
\\
A_{\mu} = (\phi , -{\bf A}) , \;
j_{\mu} = (c\rho , -{\bf j}) .
\end{eqnarray*}
Spatial components have reversed signs in the corresponding
contravariant 4-vectors, indicated by $x^{\mu}$, etc.
The Dirac matrices are represented in a form appropriate to a
2-component fermion theory, in which chirality $\gamma^5$
is diagonal for mass-zero fermions,
\begin{eqnarray}\label{Eq04}
\gamma^{\mu} =\left(\begin{array}{cc}
     0&I\\
     I&0 \end{array}\right),
\left(\begin{array}{cc}
     0& \sigv\\
  -\sigv&0 \end{array}\right);\hspace*{0.1in}
\gamma^5 =\left(\begin{array}{cc} -I&0\\ 0&I \end{array}\right).
\end{eqnarray}
\par The dynamical postulate implies covariant field equations,
\begin{equation}\label{Eq05}
\partial^\mu F_{\mu\nu} = (4\pi/c) j_{\nu}
  = (4\pi /c) (-e c\psi^{\dagger}\gamma^0\gamma_{\nu}\psi),
\end{equation}
\begin{equation}\label{Eq06}
i\hbar c\gamma ^{\mu} D_{\mu}\psi = 0  .
\end{equation}
\par In units such that $\hbar=c=1$, decomposing
$A_\mu=A^{int}_\mu+A^{ext}_\mu$ into self-interaction and external
subfields, and using notation $\Astr(x)= \gamma^{\mu}A_{\mu}(x)$,
the fermion field equation is
\begin{equation}\label{Dirmop}
\{i\gamma^\mu\partial_\mu+e\Astr^{ext}\}\psi=
  -e\Astr^{int}\psi={\hat m}\psi,
\end{equation}
defining a mass operator ${\hat m}=-e\Astr^{int}$.

\section{Self-interaction  mass as an eigenvalue}
\par The self-interaction mass operator ${\hat m} =-e\Astr^{int}$ 
acts in a Fock space defined by products of creation
and annihilation operators.  The mass operator retains the triplet-odd 
character of the classical field $A_{\mu}(x)$.  What is proposed here is
to rewrite Eq.(\ref{Dirmop}) as a renormalized Dirac equation 
\begin{equation}\label{DirA}
\{i\gamma^\mu\partial_\mu+e\Astr^{ext}-m\}\psi=
 \{{\hat m}-m\}\psi =0.
\end{equation} 
\par Eq.(\ref{DirA}) 
exhibits the algebraic structure implied by renormalization.
A canonical transformation of field operators and the vacuum state
diagonalizes the mass operator and determines a c-number mass.
This transformation mixes field components of positive and
negative energy, breaking chiral symmetry.  The resulting Dirac equation
combines chiral massless Pauli spinors into 4-component Dirac bispinors.
\par The relevant Fock space is parametrized by mass parameter
$m$ in the renormalized Dirac equation.  Since $m_0=0$ for bare 
fermions, the computed eigenvalue can be identified with 
$\delta m(m)$ in standard perturbation theory.  A consistency
condition $\delta m(m)=m$ is imposed because the right-hand
member of Eq.(\ref{DirA}) must vanish, 
\par The self-energy of a free electron can be evaluated to order
$e^2$ using Feynman's rules\cite{FEY49,PAS95}.  The relevant Feynman
diagram describes virtual emission and reabsorption of a photon of
4-momentum $k$ by a free electron. 
The present analysis replaces perturbation theory for this
virtual process by solution of the algebraic eigenvalue equation
$\{{\hat m}-m\}\psi=0$.  For restricted $k$, the electron mass $m$
should be computable to relative accuracy $\alpha=\frac{e^2}{\hbar c}$
using only single-photon virtual excitations.
It will be assumed here that the electromagnetic field
and the electronic charge have been renormalized.
\par The Fock space relevant here is characterized by a bare fermion
field $\psi(x)$ with parametric mass $m$.  Expanded in a complete
set of solutions $u_p(x)$ of the Dirac equation with mass $m$,
$\psi(x)=\sum_p u_p(x) a_p$, where $x=({\bf x},t)$. 
The amplitude coefficients $a_p$ are fermion destruction operators,
with anticommutators
\begin{equation}\label{Eq09}
\{a_p , a^{\dagger}_q \}_+ = \delta_{pq} ,
\; \{ a_p , a_q \}_+ = 0 ,
\; \{ a^{\dagger}_p , a^{\dagger}_q \}_+ = 0 .
\end{equation}
A model vacuum state is defined such that $a_p|vac\rangle=0$, where
$a_p^\dag=a_a^\dag (\epsilon_a>0),
 a_p^\dag=a_i      (\epsilon_i\leq 0)$.
\par For the interacting system considered in QED, the renormalized
Maxwell field is expanded in terms of bosonic amplitude operators as
$A_\mu(x)=\sum_k A_{\mu k}(x) b_k$, such that $b_k|vac\rangle =0$.
The physical electronic charge $-e$ defines a renormalized coupling
constant that incorporates the effects of vacuum polarization.
\par The mass operator is to be diagonalized in a basis of virtual
excitations of the model vacuum state. This replaces the bare creation
operator $a^\dag_p$ by a dressed operator whose leading terms are 
$\eta^\dag_p=a^\dag_pc_p+\sum'_k b^\dag_k a^\dag_{p-k}c_{p-k}$.
Here $a^\dag_p$ creates a bare electron of 4-momentum $p$ and
$b^\dag_k a^\dag_{p-k}$ creates both a bare electron of 4-momentum $p-k$
and a photon of 4-momentum $k$.  At this level of approximation,
the implied mass $m$ is a c-number sum over $k$ if 
the sum converges.  As in standard renormalized QED,
logarithmic divergence is indicated, and a cutoff can be introduced
consistent with the observed mass.
\par If only terms of order $\alpha$ are retained in all matrix 
elements of ${\hat m}$, the mass eigenvalue here must agree to 
this order with the Feynman self-energy\cite{FEY49,PAS95}
\begin{eqnarray}
\delta m(m)=
\frac{3\alpha}{4\pi}m\left(\ln\frac{\Lambda^2}{m^2}+\half\right),
\end{eqnarray}
introducing cutoff $\Lambda$ to eliminate a logarithmic ultraviolet
divergence. $\delta m$ vanishes for $m\to 0$. The consistency condition 
for the physical electron mass requires $\Lambda/m=\exp(286.7)$,
far beyond the range of any feasible experiment.
\par Chirality considerations imply that $\delta m(0)=0$\cite{PAS95},
consistent with the perturbation formula.  The algebra of $\gamma$
matrices implies that chirality is conserved by transition matrix
elements of the form $\int{\bar\psi}_a \gamma^\mu W^k_\mu \psi_b$.
When $m=0$, bare states of opposite energy have definite but opposite 
chirality, and cannot be mixed by such virtual transitions.
Hence virtual transitions cannot affect the mass operator.

\section{U(2) local gauge symmetry}
Gauge symmetry of classical U(2) vector fields can be analyzed following
the logic of Yang and Mills for SU(2)\cite{YAM54}.  The derivation here
follows Sect.10.3 of reference\cite{NES03}.  Elementary 2x2 matrices
are defined by the unit matrix $\tau_0$ and the Pauli matrices  
$\tauv=\{\sigma_x,\sigma_y,\sigma_z\}$.  Any element of U(2) takes the
form $\exp(i\tau_k\alpha^k)$ for real $\alpha^k(x)$\cite{CAG98}, 
using the summation convention for repeated index $k=0,1,2,3$.
The gauge field $W^k_\mu=\{B_{\mu};{\bf W}_\mu\}$ has 
U(1) and SU(2) component subfields as indicated. 
Metric tensor $g_{\mu\nu}$ here has diagonal elements $1;-1,-1,-1$. 
\par 
Covariant derivative
$D_\mu=\partial_\mu+\frac{ig}{2\hbar c}\tau_k W^k_\mu$
defines the field tensor
$W^k_{\mu\nu}=D_\mu W^k_\nu-D_\nu W^k_\mu$.
In detail,
$W^k_{\mu\nu}=\{\partial_\mu B_\nu-\partial_\nu B_\mu;
\partial_\mu{\bf W}_\nu-\partial_\nu{\bf W}_\mu
-\frac{g}{\hbar c}{\bf W}_\mu\times{\bf W}_\nu\}$.
$U(x)=I-\half ig \tau_k \chi^k(x) + \cdots$ 
determines an infinitesimal local gauge transformation
$W^k_\mu(x)\to UW^k_\mu U^{-1}+\frac{2i}{g}(\partial_\mu U(x))U^{-1}$
and implies $W^k_{\mu\nu} \to UW^k_{\mu\nu}U^{-1}$.
For U(2) fields with no kinematic mass, the Lagrangian density
${\cal L}_W=-\frth W^{\mu\nu}_k W_{\mu\nu}^k$ is gauge 
invariant\cite{REN90,CAG98,NES03}.  Conventional field strength units
here absorb a factor $4\pi$.  
\par The Euler-Lagrange equations for the gauge fields follow from
$\partial_\mu\frac{\partial{\cal L}}{\partial W^{k'}_{\tau\sigma}}      
\frac{\partial W^{k'}_{\tau\sigma}}{\partial(\partial_\mu W^k_\nu)}=
\frac{\partial{\cal L}}{\partial W^{k'}_{\tau\sigma}}
\frac{\partial W^{k'}_{\tau\sigma}}{\partial W^k_\nu}$.
The individual terms here are
\begin{eqnarray}
\frac{\partial{\cal L}}{\partial W^{k'}_{\tau\sigma}}
\frac{\partial W^{k'}_{\tau\sigma}}{\partial(\partial_\mu W^k_\nu)}&=&
-\{B^{\mu\nu};{\bf W}^{\mu\nu}\} \nonumber \\
\frac{\partial{\cal L}}{\partial W^{k'}_{\tau\sigma}}
\frac{\partial W^{k'}_{\tau\sigma}}{\partial W^k_\nu}&=&
-\frac{g}{\hbar c}\{0;{\bf W}_\mu\times{\bf W}^{\mu\nu}\},
\end{eqnarray}
implying gauge field equations $\partial_\mu B^{\mu\nu}=0$ and \\
$\partial_\mu{\bf W}^{\mu\nu}=
 \frac{g}{\hbar c}{\bf W}_\mu\times{\bf W}^{\mu\nu}=
\frac{1}{c}{\bf J}^\nu_W$.
This SU(2) self-interaction implies finite
mass for ${\bf W}_\mu$, which breaks U(2) symmetry.
$B_\mu$ has no self-interaction, and hence no resulting mass. It could
be identified with the Maxwell field $A_\mu$, except that U(2) theory
has no interaction between the U(1) and SU(2) subfields, not defining
an electromagnetic current density for the fields 
$W^{\pm}_\mu=(W^1_\mu\mp W^2_\mu)/\sqrt{2}$, which are observed to
carry unit charge.

\section{The SU(2) scalar field}
Although U(2) theory does not require a scalar Higgs field, a 
2-component scalar field $\Phi$ can be introduced without violating any 
symmetry. The covariant derivative $D_\mu\Phi$ defines a gauge-invariant
Lagrangian density ${\cal L}_\Phi=\half(D_\mu\Phi)^\dag(D^\mu\Phi)$,
with no kinetic mass term. Because ${\cal L}_\Phi$ is quadratic in the 
covariant derivatives, it also describes interaction among the gauge 
fields. The Euler-Lagrange equation for $\Phi$ takes the form
\begin{eqnarray}
\partial_\mu\frac{\partial(D_\mu\Phi)^\dag}
                  {\partial(\partial_\mu\Phi^\dag)}
 \frac{\partial{\cal L}_\Phi}{\partial(D_\mu\Phi)^\dag}&-&
 \frac{\partial(D_\mu\Phi)^\dag}{\partial\Phi^\dag}
 \frac{\partial{\cal L}_\Phi}{\partial(D_\mu\Phi)^\dag}
\nonumber \\ 
&=& D^\dag_\mu D^\mu\Phi=0, 
\end{eqnarray}
which can be written as
$\{\partial_\mu\partial^\mu+{\hat m}^2\}\Phi=0$.
\par The mass operator ${\hat m}^2$ is constructed from
gauge fields generated by virtual excitations of $\Phi$, which determine
a source current density for the gauge fields.  This implies a
self-interaction mechanism analogous to that of leptons and quarks.
The mass operator is diagonalized by a canonical transformation that
associates a minimal mass eigenvalue with a global stationary state of
the renormalized vacuum.  This determines different masses for the
renormalized component fields $\phi_0$ and $\phi_+$, breaking SU(2)
symmetry without a biquadratic term in the Lagrangian 
density\cite{CAW73}.  The transformation minimizes self-interaction 
mass by decoupling $A_\mu$, the most strongly interacting component of 
the transformed U(2) gauge field, from the transformed scalar $\phi_0$. 
As a result, $\phi_0$ has zero electric charge, and its self-interaction
mass is due only to the renormalized weak gauge fields. At the same time, 
the interaction strength of $A_\mu$ is maximized by forcing its mass 
to zero.  This eliminates the largest self-interaction mass term for both 
$\phi_0$ and $A_\mu$.  This transformation can be identified with the 
canonical transformation assumed in electroweak theory\cite{WEI96}.
\par
Assuming isospin $t=\half$ and hypercharge $y=1$, the self-interaction 
eigenstates define isospin $t_3=\mp\half$ and electric charge factor 
$Q=0,1$, which vanishes for renormalized field $\phi_0(x)$. 
A bosonic field of positive mass can vanish exactly, since 
there is nothing to exclude occupation numbers $n=0$ in physical states.
Only the component field of lowest mass would remain in
a physical ground state.  The Higgs condition $\phi_+\equiv 0$
might simply be a dynamical implication of self-interaction, restated
as $(\phi_+/\phi_0)\to 0$ for physically accessible states. 
\par For renormalized coupling constants $g_1\neq g_2$,
interactions with the gauge boson fields
arise from terms in the covariant derivative
$D_\mu\Phi=\{\partial_\mu+\frac{ig_1}{2\hbar c}B_\mu
  +\frac{ig_2}{2\hbar c} \tauv\cdot{\bf W}_\mu\}\Phi$.
If $(\phi_+/\phi_0)\to 0$,
the residual covariant derivative is\cite{WEI96} 
\begin{eqnarray}
D_\mu\Phi=\left( \begin{array}{c}
\frac{ig_2}{2\hbar c}\sqrt{2}W^+_\mu\phi_0 \\
(\partial_\mu-\frac{i}{2\hbar c}\sqrt{g_1^2+g_2^2} Z_\mu)\phi_0
\end{array} \right),
\end{eqnarray}
where $Z_\mu=W^3_\mu\cos\theta_W-B_\mu\sin\theta_W$ is 
a linear transform of the neutral gauge fields. 
The Weinberg angle $\theta_W$ is defined such that
$\sin\theta_W=g_1/\sqrt{g_1^2+g_2^2}$.   The decoupled orthogonal field 
$A_\mu=W^3_\mu\sin\theta_W+B_\mu\cos\theta_W$
is the physical Maxwell field.
The Lagrangian term $-\half W_-^{\mu\nu}W^+_{\mu\nu}$ is replaced by
$-\half{\tilde W}_-^{\mu\nu}{\tilde W}^+_{\mu\nu}$, 
omitting interaction terms among the weak gauge fields,
where 
\begin{eqnarray}
{\tilde W}^+_{\mu\nu}&=& 
 (\partial_\mu+\frac{ig_2}{\hbar c}\sin\theta_W A_\mu)W^+_\nu
\nonumber \\
&-&(\partial_\nu+\frac{ig_2}{\hbar c}\sin\theta_W A_\nu)W^+_\mu.
\end{eqnarray}
This transformed Lagrangian exhibits U(1) gauge invariance for $Q=+1$.
The implied electric charge unit is
$e=g_2\sin\theta_W=-g_1\cos\theta_W$\cite{WEI96,CAG98}.
In the appropriate energy range\cite{WEI96}, 
$e=\sqrt{4\pi/129}=0.3121$
in units such that $\hbar=c=1$.  The Fermi constant
$G_F=1.16639\times 10^{-5}$(GeV)$^{-2}$ determines the ratio
$(g_2/M_W)^2\simeq4\sqrt{2}G_F$.
Combining these values\cite{WEI96}, the empirical mass 
$M_W=$ 80.33 GeV implies coupling constants $g_1=$0.3554, $g_2=$0.6525, 
so that $\sin^2 \theta_W=$0.2288 (accepted value 0.2315)\cite{CAG98}.
\par Denoting SU(2) self-interaction mass, without the scalar field,
by $M_0$, $\Delta M_W= M_W-M_0$ and $\Delta M_Z= M_Z-M_0$ such that
$\cos\theta_W=\Delta M_W/\Delta M_Z$ for the symmetry-breaking mass
induced by the scalar field.  Empirical $M_W=$80.33 Gev, 
$M_Z=$91.19 Gev, and $\sin^2 \theta_W=$0.2315 imply $M_0=$3.15 GeV
and $\Delta M_W=$77.18 GeV, $\Delta M_Z=$88.04 GeV.
\par The transformed Lagrangian density for $\phi_0$ is
\begin{eqnarray}
{\cal L}_\phi=
\half\partial_\mu\phi_0 \partial^\mu\phi_0 &+&
\nonumber \\
 \frth\frac{g_2^2}{(\hbar c)^2}W^-_\mu W_+^\mu\phi_0^2
&+&\egth\frac{g_1^2+g_2^2}{(\hbar c)^2}Z_\mu Z^\mu\phi_0^2. 
\end{eqnarray}
Because $\phi_0$ interacts only with the weak gauge fields, it is
electrically neutral.  It has no electromagnetic self-interaction 
or resulting mass, while the Maxwell field remains massless. 
The interaction terms
\begin{eqnarray}
\frac{\partial{\cal L}_\phi}{\partial W^-_\mu}=
 \frth\frac{g_2^2}{(\hbar c)^2}W_+^\mu\phi_0^2 ;\;\; 
 \frac{\partial{\cal L}_\phi}{\partial Z_\mu}=
 \frth\frac{g_1^2+g_2^2}{(\hbar c)^2} Z^\mu\phi_0^2  
\end{eqnarray}
define dynamical mass in the gauge field equations\cite{WEI96}.  These 
terms imply mass proportional to $|\phi_0|$ for both $W^\pm_\mu$ and
$Z_\mu$.  They augment the direct self-interaction in the SU(2) field 
equations and break SU(2) symmetry.
\par Any biquadratic term in the Lagrangian implies a nonlinear
Poisson-like field equation for $\phi_0$.  Yukawa coupling to fermions
implies an inhomogeneous equation driven by fermion densities. 
The signs and magnitudes of these terms may 
have cosmological implications\cite{MAN06}.

\section{Conclusions}
If the unique source of mass in electroweak theory is direct or
indirect self-interaction, traditional quantum field theory 
contains formal mechanisms that for fermions can explain
the finite but small mass of neutrinos, while providing a rationale
for the existence of fermion generations distinguished only by mass.
Applied to a scalar (Higgs) boson in a nominal U(2) manifold,
symmetry-breaking consistent with the electroweak model is driven 
by a canonical transformation that minimizes the scalar mass 
and forces the residual Maxwell field to be massless.  The present 
analysis implies that the self-interaction mass of the Higgs boson 
may arise from weak interactions only, and might be very small,
analogous to neutrino mass.  This is the only physical implication
contrary to the Standard Model, and is clearly subject to experimental
test.  It is inconsistent to postulate that neutrino mass is exactly
zero, since self-interaction mediated by weak gauge fields is
implied\cite{NES03}.
   

\end{document}